\pdfoutput=1
\documentclass[conference,letterpaper]{IEEEtran}

\usepackage[noadjust]{cite}  %
\usepackage{balance}  %
\usepackage{url}  %
\usepackage{colortbl}  %
\definecolor{Gray}{gray}{0.94}
\usepackage{graphicx}  %

\definecolor{newgreen}{RGB}{7, 160, 25}
\definecolor{newblue}{RGB}{100, 100, 200}
\definecolor{newpurple}{RGB}{200, 100, 200}

\newenvironment{gn}{\color{newgreen}}

\newenvironment{bl}{\color{newblue}}

\newenvironment{pr}{\color{newpurple}}

\newcommand{\docauthor}{Jonathan Will\IEEEauthorrefmark{1}, Nico Treide\IEEEauthorrefmark{1}, Lauritz Thamsen\IEEEauthorrefmark{2}, and Odej Kao\IEEEauthorrefmark{1}}
\newcommand{\docsubject}{\IEEEauthorrefmark{1}Technische Universit\"at Berlin, Germany \hspace{6mm} \IEEEauthorrefmark{2}University of Glasgow, United Kingdom}
\newcommand{\dockeywords}{Scalable Data Analytics, Distributed Dataflows, Resource Allocation, Autoscaling, Cluster Management}
\newcommand{\doctitle}{Experimentally Evaluating the Resource Efficiency of Big Data Autoscaling}

\PassOptionsToPackage{bookmarks=false}{hyperref}
\usepackage[pdftex]{hyperref}
\hypersetup{
    pdfborder={0 0 0},
    pdfauthor=\docauthor,
    pdftitle=\doctitle,
    pdfsubject=\docsubject,
    pdfkeywords=\dockeywords
}

\def\mycopyrightnotice{
  {\footnotesize 978-1-6654-9115-0/22/\$31.00~\copyright~2024 IEEE\\
   DOI: \href{https://doi.org/10.1109/BigData62323.2024.10825367}{https://doi.org/10.1109/BigData62323.2024.10825367}
  }
  \gdef\mycopyrightnotice{}
}
\makeatletter
  \def\ps@IEEEtitlepagestyle{%
 \def\@oddfoot{\mycopyrightnotice}
  \def\@evenfoot{}
  }%
\makeatother

\def\BibTeX{{\rm B\kern-.05em{\sc i\kern-.025em b}\kern-.08em T\kern-.1667em\lower.7ex\hbox{E}\kern-.125emX}}

\begin{document}

\title{\doctitle}

\author{%
\IEEEauthorblockN{\docauthor}
\IEEEauthorblockA{\docsubject\\
\{will, nico.treide, odej.kao\}@tu-berlin.de \hspace{6mm} lauritz.thamsen@glasgow.ac.uk
}}

\maketitle

\begin{abstract}
Distributed dataflow systems like Spark and Flink enable data-parallel processing of large datasets on clusters.
Yet, selecting appropriate computational resources for dataflow jobs is often challenging.
For efficient execution, individual resource allocations, such as memory and CPU cores, must meet the specific resource requirements of the job.
An alternative to selecting a static resource allocation for a job execution is autoscaling as implemented for example by Spark.

In this paper, we evaluate the resource efficiency of autoscaling batch data processing jobs based on resource demand both conceptually and experimentally by analyzing a new dataset of Spark job executions on Google Dataproc Serverless.
In our experimental evaluation, we show that there is no significant resource efficiency gain over static resource allocations.
We found that the inherent conceptual limitations of such autoscaling approaches are the inelasticity of node size as well as the inelasticity of the ratio of memory to CPU cores.

\end{abstract}

\IEEEpeerreviewmaketitle


\section{Introduction}

Large-scale batch data processing has diverse application areas such as science and commerce.
Distributed dataflow systems like Spark~\cite{spark} and Flink~\cite{flink} simplify developing scalable data-parallel programs, reducing the need to implement parallelism and fault tolerance while using clusters of commodity resources.
Major cloud providers offer dedicated services such as Amazon EMR or Google Dataproc, allowing users to deploy their jobs\footnote{By job, we mean a data processing algorithm, implemented in a specific system, and running on a given input dataset. In Spark's terminology, this would be an application.} to a cluster.

Yet, configuring a suitable cloud cluster for a given job is still difficult~\cite{lama2012aroma,rajan2016perforator}.
At a minimum, it involves selecting the number of nodes, the number of CPU cores per node, and the amount of memory per node, resulting in many possible options.
Overprovisioning resources can lead to low resource utilization, unnecessarily increasing cost\footnote{``Cost'' can manifest as, e.g., monetary cost, capacity consumption, or carbon emissions.}~\cite{yang2013bubble,liu2011measurement,delimitrou2014quasar,lin2013scaling}.
Meanwhile, underprovisioning of one type of resource can lead to resource bottlenecks, causing the ensemble of underperforming cluster resources to incur more costs by being occupied for a longer period of time~\cite{al2022juggler,al2022blink,will2022get,will2023selecting}.
Conversely, proper resource allocation simultaneously optimizes both cost and performance by allocating resources that are best suited for the given workload.
However, performance itself is often not a major concern in non-interactive data analytics, e.g., in data analytics jobs running over night.
Therefore, our primary focus is to address the problem from the perspective of optimizing resource efficiency and only secondarily from the perspective of optimizing performance.

There are two main types of approaches to solving the problem of suitable resource allocation: Static resource allocation and dynamic resource allocation.

\emph{Static resource allocation} strategies select a given set of cluster resources to use throughout the entire job execution~\cite{ernest,cherrypick,hsu2018arrow}.
Here, a suitable allocation is typically estimated by performance models based on historical executions of similar jobs.

\emph{Dynamic resource allocation} strategies start with a typically small default configuration of cluster resources and continuously adjust this allocation during the job execution~\cite{scheinert2021enel}.
Reasons for reallocation may include reacting to different resource needs of different phases of a data processing job, reacting to changing resource availablility (job priority, resource prices, green energy availablility, etc.), or trying to complete the job within a given deadline.
Spark implements an autoscaling approach based on handling fluctuating resource demand and cloud providers like Google Dataproc offer this autoscaling to users as a way to avoid having to manually configure resources.

In this paper, we evaluate the resource efficiency of autoscaling batch data processing jobs based on resource demand both conceptually and experimentally.
We compare Google Dataproc Serverless, which is a managed Spark service offering autoscaling to a baseline of static resource allocations by analyzing a new dataset of Spark job executions.
Based on our observations, we then discuss the design limitations of such an autoscaling method with regards to optimizing resource allocations for efficiency.

\vspace{3mm}
\hspace{-5mm}
\emph{Contributions.} The contributions of the paper are as follows.

\begin{itemize}
    \item A new benchmarking suite called \emph{benchspark}\footnote{\href{https://github.com/dos-group/benchspark}{github.com/dos-group/benchspark}}, which contains a diverse set of Spark jobs
    \item A new trace dataset\footnote{\href{https://github.com/dos-group/spark-autoscaling-evaluation}{github.com/dos-group/spark-autoscaling-evaluation}} of executions of said Spark jobs on ``Dataproc Serverless'' on Google Cloud Platform
    \item An experimental evaluation of Spark job autoscaling on Dataproc against a baseline of static resource allocations
    \item A discussion about the conceptual viablility of Spark's autoscaling approach for resource efficiency
\end{itemize}

\section{Background}

This section explains several concepts that are fundamental to automatically scaling batch data processing jobs.

\subsection{Resource Allocation for Distributed Dataflow Jobs}

Distributed dataflow jobs use different types of resources for their execution, such as memory, CPU, network, and storage.
The latter two are not always configurable in every infrastructure, while memory and CPU, as well as the number of nodes in the cluster, are typically configurable.
To select an efficient configuration of resources, one must consider the resource access patterns of a given job, as well as the current cost (e.g., availablility or price) of individual resources such as memory and CPU.
One crucial influence on the resource usage patterns of a dataflow job is the underlying algorithm and its implementation in a distributed dataflow system.

For instance algorithms that iterate over a dataset only once typically require only small memory allocations, while any more memory would only increase the cost of execution without any tangible benefit.
Other jobs may benefit from holding at least a part of the input dataset in memory to avoid slow disk reads.
This can result in a lower runtime and thereby lower execution cost through means of occupying the ensemble of resources for a shorter period of time, as long as the cost of memory is not too high.
Similarly, there are jobs that are highly parallelizable and can therefore benefit from more CPU cores in the cluster, while for other jobs this results in additional cost and little performance gain.
Once the total number of CPU cores and the amount of memory in the cluster have been configured, one has the option of distributing these resources across many small nodes or fewer but larger nodes.
The former option can lead to increased network traffic, while the latter option can lead to disk bottlenecks, depending on the particular job at hand.

\subsection{Autoscaling Resources for Distributed Dataflow Jobs}

Apache Spark is a prominent distributed data processing engine for general purpose large-scale data analytics.
To address fluctuating resource demands of data processing jobs, Spark introduced \emph{dynamic resource allocation} as a mechanism to optimize resource usage by automatically adjusting the number of \emph{executors} during runtime based on workload demands.
This capability allows Spark to allocate more resources when needed and to free up idle executors when workloads decrease.
It uses external shuffle services to ensure that data is not lost along with nodes that are being removed during downscaling.

However, Spark's dynamic resource allocation is limited by the constraints of the underlying cluster infrastructure.
In a traditional Spark deployment, cluster resources (such as virtual machines or physical nodes) are often statically provisioned by the user or a cluster manager like \emph{YARN} or \emph{Kubernetes}.
While Spark can dynamically scale the number of executors, it cannot autonomously scale the underlying physical infrastructure, such as adding or removing nodes based on job requirements.
As a result, Spark's dynamic allocation is constrained by the pre-provisioned cluster capacity, limiting its ability to fully optimize for resource usage and cost efficiency in environments where workloads fluctuate significantly.

On public cloud platforms, managed Spark services that offer resource allocation via autoscaling, such as Azure HDInsight, Amazon EMR, and Google Dataproc, are available as an alternative to making users select a static allocation for the duration of their job.
These autoscaling offerings extend Spark's dynamic allocation by also managing the \emph{underlying cluster infrastructure}.
This allows them to automatically scale up and down both the number of executors and the associated virtual machines dynamically allocated to the job.
Users can set a minimum and maximum allocation of executors, i.e., worker nodes.

These extensions to the original Spark autoscaling are more suitable for modern cloud-based data processing environments, where infrastructure can be provisioned elastically, and the cost of underutilized resources is a primary concern.
By dynamically adjusting both executors and nodes, managed Spark services' autoscaling aims to achieve a more efficient use of resources compared to static resource provisioning or standalone Spark.

\section{Experimental Setup}

To empirically evaluate different resource allocation options, including autoscaling, we ran a set of diverse Spark jobs on cloud configurations with varying scale-out, memory per node, and CPU per node, and recorded the resulting runtimes.
In this section we describe this trace dataset.

\subsection{Spark Jobs}

As depicted in Table~\ref{tab:jobs}, we created 18 test jobs from nine common underlying data processing algorithms and two differently sized input datasets for each algorithm.
These jobs were compiled with Scala 2.12.14 and ran on Spark 3.3.0, using Java 11.

\begin{table}[htb]
  \center
  \caption{The 18 Spark Jobs Executed on Google Dataproc}
  \begin{tabular}{lll}
    Algorithm & Data Type & Dataset Sizes [GiB] \\
    \hline\\[-2ex]
    Grep & Text & \{ 3010, 6020 \} \\\rowcolor{Gray}
    Sort & Text & \{ 94, 188 \} \\
    Word Count & Text & \{ 39, 77 \} \\\rowcolor{Gray}
    K-Means & Vector & \{ 102, 204 \} \\
    Linear Regression & Vector & \{ 229, 459 \} \\\rowcolor{Gray}
    Logistic Regression & Vector & \{ 210, 420 \} \\
    Join & Tabular & \{ 85, 172 \} \\\rowcolor{Gray}
    GroupByCount & Tabular & \{ 280, 560 \} \\
    SelectWhereOrderBy & Tabular & \{ 92, 185 \} \\
  \end{tabular}
  \label{tab:jobs}
\end{table}

The jobs consist of commonly known data processing algorithms, for which we used the standard implementations available in Spark's libraries.
The actual source code of the jobs and the test dataset generators can be found in a public git repository$^3$.
While these types of jobs have also been used in the creation of older such trace datasets~\cite{hsu2018arrow,will2020towards}, our new dataset aims to weigh the jobs targeting different data types equally, in an attempt to create a balanced sample.
Specifically, these are text data, vector data, and tabular data.

\subsection{Cloud Configurations}

Table~\ref{tab:serverless_configurations} lists the three different Dataproc Serverless configurations we used to execute each of the 18 jobs, resulting in a total of 54 job executions.
For reference, we assigned an ID to each of the Serverless configurations.
Configuration \texttt{S1} has a minimum and maximum number of executors of 2 and 32 respectively and uses autoscaling.
Configurations \texttt{S2} and \texttt{S3} remain static at 8 and 16 nodes respectively and serve as a baseline against which to evaluate the dynamically scaling \texttt{S1}.

\begin{table}[b]
  \center
  \caption{Cloud Configurations Used for Job Execution\\with Serverless Google Dataproc}
  \begin{tabular}{rlccr}
    ID & Instance Type & Scale-Out & Total Cores & Total RAM \\
    \hline\\[-2ex]
    \texttt{S1} & n2-standard-4 & 2-32 & 8-128 & 32-512 GiB \\\rowcolor{Gray}
    \texttt{S2} & n2-standard-4 & 8-8 & 32-32 & 128-128 GiB \\
    \texttt{S3} & n2-standard-4 & 16-16 & 64-64 & 256-256 GiB \\
  \end{tabular}
  \label{tab:serverless_configurations}
\end{table}

Table~\ref{tab:regular_configurations} lists the ten additional GCP configurations we used to execute each of the 18 jobs, resulting in an additional 180 job executions.
These executions serve to further put the results of \texttt{S1}, \texttt{S2}, and \texttt{S3} into perspective.

Configurations \texttt{R01} through \texttt{R03} differ only in total cluster memory, while configurations \texttt{R04} through \texttt{R06} differ only in total cluster CPU cores.
The remaining configurations share total memory and total CPU with at least one other configuration and only differ in scale-out.
Thus, our choice of cloud configuration space and the resulting runtime dataset also enables isolating and interpreting each of these three influences’ impact on a given job’s runtime.
Further, the regular Dataproc configurations \texttt{R08} and \texttt{R09} aim to replicate the Dataproc Serverless configurations \texttt{S2} and \texttt{S3} respectively.

Overall, the configuration options in our evaluation dataset do not focus solely on scale, as this would mostly translate into a cost/performance tradeoff~\cite{will2023selecting}.
Instead, the prominent configuration dimensions include the ratio of memory to CPU cores, as well as the distribution of these given resources across fewer large nodes or more numerous but smaller nodes.
This in turn creates a search space of configuration options with different degrees of \textit{efficiency}, i.e., distance from the cost-performance Pareto front.

\vspace{3pt}

\begin{table}[b]
  \center
  \caption{Cloud Configurations Used for Job Execution\\with Regular Google Dataproc}
  \begin{tabular}{rlrrr}
    ID & Instance Type & Scale-Out & Total Cores & Total RAM \\
    \hline\\[-2ex]
    \texttt{R01} & n2-highcpu-8 & 8 & 64 & 64 GiB      \\\rowcolor{Gray}
    \texttt{R02} & n2-standard-8 & 8 & 64 & 256 GiB    \\
    \texttt{R03} & n2-highmem-8 & 8 & 64 & 512 GiB     \\\rowcolor{Gray}
    \texttt{R04} & n2-highmem-4 & 4 & 16 & 128 GiB     \\
    \texttt{R05} & n2-standard-8 & 4 & 32 & 128 GiB    \\\rowcolor{Gray}
    \texttt{R06} & n2-highcpu-32 & 4 & 128 & 128 GiB   \\
    \texttt{R07} & n2-highmem-8 & 2 & 16 & 128 GiB     \\\rowcolor{Gray}
    \texttt{R08} & n2-standard-4 & 8 & 32 & 128 GiB    \\
    \texttt{R09} & n2-standard-4 & 16 & 64 & 256 GiB   \\\rowcolor{Gray}
    \texttt{R10} & n2-highcpu-8 & 16 & 128 & 128 GiB   \\
  \end{tabular}
  \label{tab:regular_configurations}
\end{table}

\subsection{Job Executions}

\hspace{-3pt}
Table~\ref{tab:executions} shows the statistical properties of the trace dataset that resulted from executing each of the 18 Spark jobs on each of the 13 resource configurations.

\vspace{3pt}

\begin{table}[htb]
  \center
  \caption{Statistical Properties of the Evaluation Trace Dataset Containing 234 Spark Job Executions}
  \begin{tabular}{lrrr}
      & Runtime [Seconds] & vCPU Core Seconds & GiB RAM Seconds \\
    \hline\\[-2ex]
    mean   &  1,663.46 &    80,739.76 &    256,184.50 \\\rowcolor{Gray}
    std.   &  2,615.11 &   176,908.69 &    332,856.28 \\
    min.   &    141.68 &     8,196.64 &     16,662.40 \\\rowcolor{Gray}
    25\%   &    466.59 &    27,257.35 &     90,759.68 \\
    50\%   &    853.85 &    43,597.20 &    171,200.12 \\\rowcolor{Gray}
    75\%   &  1,615.02 &    69,456.06 &    276,902.64 \\
    max.   & 21,714.74 & 2,038,734.08 &  2,779,486.72 \\

  \end{tabular}
  \label{tab:executions}
\end{table}

\vspace{3pt}

On average, a job execution cost about \$1.29 USD and lasted about 28 minutes.
The metrics vCPU core seconds and GiB RAM seconds represent the product of the amount and duration of usage for vCPU cores and GiB of memory respectively.

Note that due to budget constraints, each job was executed only once on each cloud resource configuration, which may make this measured test job data somewhat susceptible to individual outliers.
However, we expect the overall evaluation to be accurate as it is mostly based on averages.

\vspace{20pt}

\section{Experimental Evaluation}

In this section, we evaluate the dataset of Google Dataproc job executions.

\subsection{Scaling Behavior of Dataproc Serverless}

Google Dataproc Serverless dynamically adjusts the allocation of executors based on Spark's autoscaling feature.
As specified as configuration \texttt{S1}, we kept the default minimum number of executors (2) and set the maximum number of nodes to 32.
Besides this maximum, we kept all of the default settings.

\begin{figure}[htb]
  \includegraphics[width=\linewidth, keepaspectratio]{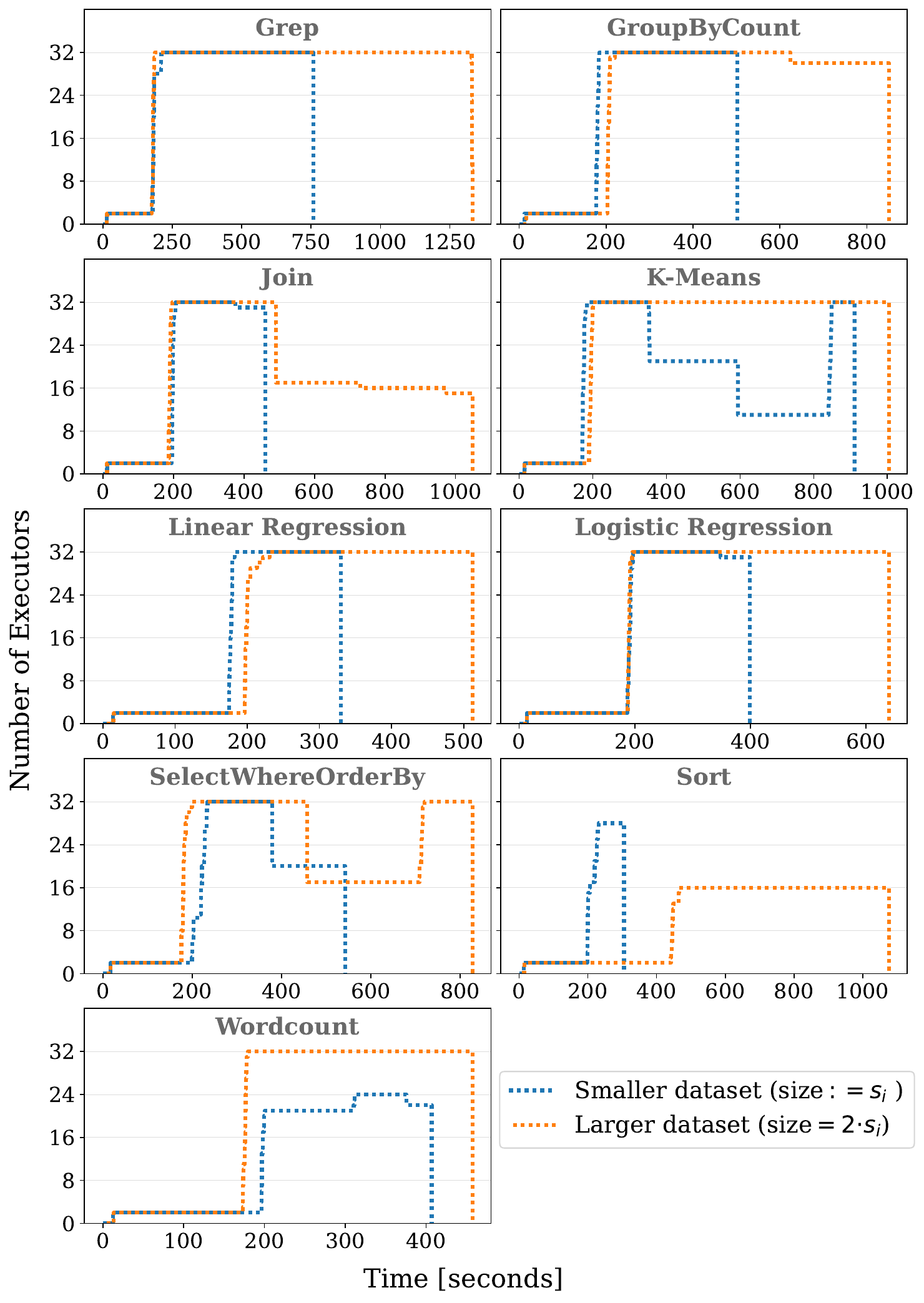}
  \caption{Time series of Dataproc Serverless executor allocation for executors in $[2,32]$.}\label{timeseries}
\end{figure}

In Figure~\ref{timeseries}, we can observe that for some jobs, there is an intensive processing phase, where the maximum number of executors are being allocated right after about 180 to 200 seconds of runtime, which appears to be the latency for the initial demand for more executors to materialize.
We see that for some jobs, processing the larger dataset takes much more than twice the amount of time, while for others, processing the larger dataset takes much less than twice the amount of time.
We also see that for some jobs, the scaling behavior appears similar for both datasets, e.g., Logistic Regression or Linear Regression.
On the other hand, for some jobs, the scaling behavior for both datasets can be dissimilar, e.g. the K-Means jobs.
\clearpage

\begin{table}[h!]
  \center
  \caption{Expended Executor Seconds\\Normalized to 1 = Lowest per Job}
  \begin{tabular}{lr|rrr}
    && \texttt{S1} & \texttt{S2} & \texttt{S3} \\
    \hline\\[-2ex]
                  Grep & Large &     1.087  & \gn{1.000} &     1.022  \\
                  Grep & Small &     1.057  & \gn{1.000} &     1.013  \\\rowcolor{Gray}
          GroupByCount & Large &     1.041  & \gn{1.000} &     1.007  \\\rowcolor{Gray}
          GroupByCount & Small &     1.095  & \gn{1.000} &     1.068  \\
                  Join & Large &     1.181  & \gn{1.000} &     1.072  \\
                  Join & Small &     1.203  & \gn{1.000} &     1.073  \\\rowcolor{Gray}
               K-Means & Large & \gn{1.000} &     1.437  &     1.019  \\\rowcolor{Gray}
               K-Means & Small &     1.247  &     1.007  & \gn{1.000} \\
      LinearRegression & Large &     1.036  &     1.039  & \gn{1.000} \\
      LinearRegression & Small &     1.126  &     1.082  & \gn{1.000} \\\rowcolor{Gray}
    LogisticRegression & Large & \gn{1.000} &     2.004  &     1.147  \\\rowcolor{Gray}
    LogisticRegression & Small &     1.023  &     1.197  & \gn{1.000} \\
    SelectWhereOrderBy & Large &     1.119  &     1.005  & \gn{1.000} \\
    SelectWhereOrderBy & Small &     1.167  & \gn{1.000} &     1.057  \\\rowcolor{Gray}
                  Sort & Large &     1.671  &     1.080  & \gn{1.000} \\\rowcolor{Gray}
                  Sort & Small &     1.105  &     1.038  & \gn{1.000} \\
             WordCount & Large &     1.101  & \gn{1.000} &     1.026  \\
             WordCount & Small &     1.163  & \gn{1.000} &     1.050  \\
  \end{tabular}
  \label{tab:efficiency}
\end{table}

\begin{figure}[htb]
  \center
  \includegraphics[width=.7\linewidth, keepaspectratio]{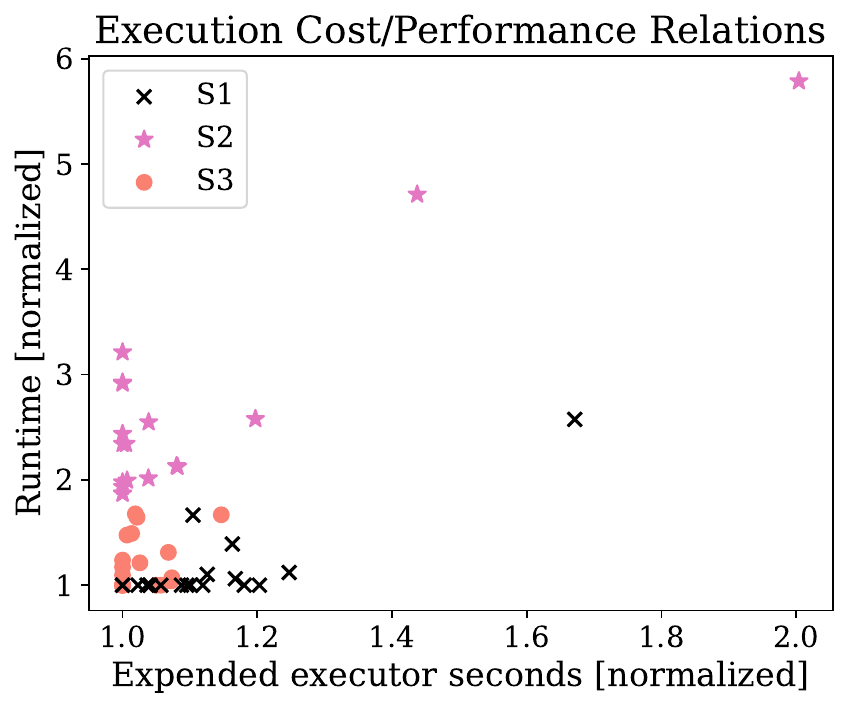}
  \caption{Comparing the achieved runtime and the resulting resource use for each of the 18 jobs when using different executor allocation strategies. Normalization to 1 = lowest per job.}\label{fig:tradeoff}
\end{figure}

\newpage
\subsection{Resource-Efficiency of Dataproc Serverless Autoscaling}

In Table~\ref{tab:efficiency}, we see a comparison of the expended executor seconds per Dataproc Serverless resource configuration, normalized so that a value of 1 represents the lowest by any of the three configurations for the given job.
This helps to highlight only the relative difference between the executions of a job with the three different configurations.
We can observe that the number of executor seconds used to run a given job was the lowest for the dynamic configuration \texttt{S1} in 2 instances, while the static allocations of 8 and 16 executors (\texttt{S2} and \texttt{S3}) achieved the lowest executor seconds 9 and 7 times respectively.
The most cost-efficient configuration option appears to largely remain similar for different jobs that share the same algorithm, and appears less related to the size of the input dataset.
The largest variance is seen for Logistic Regression on the larger dataset, where configuration \texttt{S1} executed the job with roughly half the amount of expended executor seconds compared to \texttt{S2}.
In general, however, the differences are much smaller, resulting in roughly similar levels of aggregated resource usage over time for each of these three configuration options.

Figure~\ref{fig:tradeoff} compares the resource usage along with the runtime achieved for each of the 18 jobs when using the three different Dataproc Serverless configurations.
Again, the values are normalized for each job so that a value of 1 represents the lowest value that any of the 3 configurations achieved for that job.
First, we see that the variation in runtimes is much larger than the variation in expended executor seconds.
Second, we see that the dynamically scaling configuration \texttt{S1} produces the lowest runtimes for most jobs, while the runtimes increase when using a static 16 executors and even more drastically when using 8 executors.

These results suggest, that while resource efficiency benefits of Dataproc's Spark autoscaling are rare, jobs typically finish faster with autoscaling than with static allocations.

\vspace{3mm}

\subsection{Impact of the Configuration Space Dimensions on Cost}

In general, there are more configuration parameters to set than just the scale-out.
These include in particular the node size and the resource mix per node.
To evaluate the impact of these additional configuration options on resource efficiency, we compare the results of Dataproc Serverless job executions with the same executions on the static configurations \texttt{R01} to \texttt{R10}.
Instead of using executor seconds as the cost metric, here we use GiB memory seconds and vCPU core seconds, applying a factor of 1:6, meaning that one second of using 6 GiB memory is considered to cost as much as one second of using 1 vCPU core\footnote{\href{https://cloud.google.com/dataproc-serverless/pricing}{cloud.google.com/dataproc-serverless/pricing}}.
This represents the pricing model of Google Dataproc Serverless for RAM and CPU as of October 2024, while contemporary pricing of the general purpose N2 VMs in GCP's Frankfurt datacenter has a roughly similar ratio of about 1:7.46.

\begin{table}[htb]
  \center
  \caption{Evaluating Cost and Performance for Different Resource Configurations of Regular and Serverless Dataproc.\\Values Normalized to 1 = Lowest per Job\\and then Averaged Across all 18 Jobs.}
  \begin{tabular}{l|rr}
    & Mean Normalized Cost & Mean Normalized Runtime \\
    \hline\\[-2ex]
      \texttt{R01}         &               1.875  &     3.410   \\\rowcolor{Gray}
      \texttt{R02}         &               1.348  &     1.727   \\
      \texttt{R03}         &               1.643  &     1.511   \\\rowcolor{Gray}
      \texttt{R04}         &               1.849  &     6.820   \\
      \texttt{R05}         &               1.739  &     4.462   \\\rowcolor{Gray}
      \texttt{R06}         &               5.455  &     4.889   \\
      \texttt{R07}         &               2.392  &     8.792   \\\rowcolor{Gray}
  \bl{\texttt{R08}}        &               1.334  &     3.433   \\
  \pr{\texttt{R09}}        &           \gn{1.286} &     1.654   \\\rowcolor{Gray}
      \texttt{R10}         &               1.598  & \gn{1.445}  \\
      \texttt{S1}          &               1.618  &     1.845   \\\rowcolor{Gray}
  \bl{\texttt{S2}}         &               1.545  &     4.021   \\
  \pr{\texttt{S3}}         &               1.475  &     1.934   \\
    \hline\\[-2ex]
  Mean & 1.935 & 3.534 \\
  \end{tabular}
  \label{tab:regular}
\end{table}

Table~\ref{tab:regular} shows that the configurations that executes jobs with the lowest incurred cost are \texttt{R08} and \texttt{R09}.
Configurations \texttt{R08} and \texttt{R09} correspond to configurations \texttt{S2} and \texttt{S3} respectively in terms of number of nodes, as well as vCPU cores and memory per node.
These configurations exhibit a slight performance deviation between regular Dataproc and Dataproc Serverless, about 17.6\% on average, with a standard deviation of about 18.5\%.
This could be explained, for example, by different levels of interference from other co-located applications that are present in each of these two service types.

Accounting for this inherent performance deviation, one can expect that \texttt{S1} would produce a cost that is at least not exceedingly far from the lowest cost on average out of the given configurations, if they were all running on the same platform.
Furthermore, always choosing the configuration that is best across all jobs on average results in costs that are about a quarter higher than what could be achieved by always selecting the most suitable configuration for each job individually.

\subsection{Impact of the Resource Cost Model}

However, once there is an abundance or scarcity of either memory or CPU cores at a given point, the cost of operating different types of resources changes, thereby affecting the viability of different resource configuration options for executing data processing jobs.
This fluctuation in availability could be caused by other non-permanent applications running at a given point in the cluster.

\begin{figure}[h!]
\center
  \includegraphics[width=.97\linewidth, keepaspectratio]{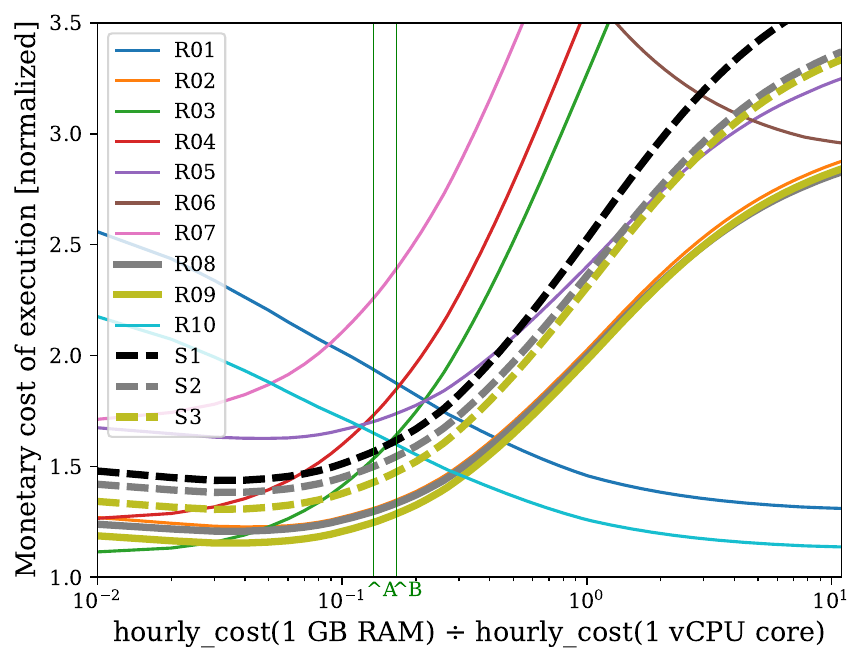}
  \caption{Comparing cloud configuration selection approaches for varying individual resource costs. Bold lines of the same color represent corresponding configurations in regular and serverless Dataproc.}\label{fig:prices_experiment}
\end{figure}

In Figure~\ref{fig:prices_experiment}, we show how the cost efficiency of different resource configuration options changes as the resource cost model changes.
On the far left of the x-axis ($10^{-2}$), the hourly cost of 1 GB of memory is equal to the hourly cost of 0.01 vCPU cores.
On the far right of the x-axis ($10^1$), the hourly cost of 1 GB of memory is equal to the hourly cost of 10 vCPU cores.
For reference, we mark the contemporary GCP VMs / Dataproc Serverless price points with a thin green vertical line and the labels A and B, respectively.

We see that some configurations exhibit lower job execution costs relative to other configurations when RAM is cheaper relative to CPU (left side of the graph) and vice versa.
This is largely a function of how much of a given resource a configuration has, and is exemplified by a comparison of the memory-heavy configuration \texttt{R03} and the CPU-heavy configuration \texttt{R10}.
It appears that Serverless Dataproc has a suitable default RAM and CPU for its executors/nodes, given the cost structure of these resources and the nature of Spark jobs.
However, in the case of a fluctuating resource cost structure, the GiB RAM to vCPU ratio for autoscaling executors would need to be adjusted in order to still reach comparatively good resource efficiency.

Note that we focus only on the \textit{relative} operating costs of these individual resources, because we are comparing only the relative cost differences of operating different resource configurations. %
Further, we disregard potential cost fluctuation of other resources, like storage space or network, since those are not part of the resource configuration space in our experiments.

\vspace{-2mm}

\newpage

\section{Discussion}

The examined Spark autoscaling approach appears to be based on selecting the empirically most optimal node.
It then continuously adjusts only the scale-out at runtime.
This is a relatively simple resource configuration attribute to adjust at runtime, as it avoids disrupting execution on the unchanged nodes.
For example, when scaling out from, say, 12 to 16 nodes, the 12 original nodes can retain cached data and continue executing their current tasks while the new nodes start up.
Similarly, when scaling in, idle nodes can be removed without interrupting the others, whereas downsizing running nodes would cause an interruption in execution.

However, our experimental evaluation has shown that adapting the node type per job has a more significant impact on resource efficiency than continuously adjusting the number of nodes in the cluster at runtime.
While the resource efficiency of autoscaling was, on average, roughly comparable to that of the historically most resource-efficient static allocation, always selecting the correct static allocation could have theoretically saved about an additional quarter of the costs.
This would involve configuring a suitable amount of memory and number of CPU cores per node in addition to selecting a suitable scale-out.
However, that requires an understanding of how each of these resources influences performance, as well as the current operating costs of these resources.
Taking these aspects into consideration is beyond the scope of today's prominent distributed dataflow systems.

The lack of adaptability to changing availability of individual resources may make the autoscaling approach more suitable for larger scale infrastructures where individual applications cause smaller variations in the mix of available resource types.

\section{Conclusion}

In this paper we have experimentally evaluated the horizontal autoscaling of Spark jobs as implemented by Dataproc on GCP from a resource efficiency perspective.
The results suggest that the cost savings are not significant when compared to some static allocations and that there is further potential for cost optimization to be realized.

In the future, we plan to continue our research into cost optimization of distributed dataflow jobs by adapting the allocation strategy for each individual job based on resource access patterns.

\section*{Acknowledgments}

This work has been supported through a grant by the German Research Foundation (DFG) as “C5” (grant 506529034) and the Google Cloud Research Credits Program.

\bibliographystyle{IEEEtran}
\balance

\end{document}